\documentclass[prd,a4paper,showpacs]{revtex4}
\usepackage{epsfig}
\usepackage{color}
\usepackage{amssymb}
\usepackage{amsfonts}
\usepackage{graphicx}
\usepackage{keyval,graphicx}
\usepackage{textcomp,wasysym}
\newcommand{\be}{  \begin{eqnarray}}
\newcommand{\ee}{\end{eqnarray}}

\begin{document}
\title{Open charm effects in $e^+e^-\to J/\psi \eta$, $J/\psi \pi^0$  and
$\phi\eta_c$}
\author{Qian Wang$^1$, Xiao-Hai Liu$^1$, and Qiang Zhao$^{1,2}$}

\affiliation{1) Institute of High Energy Physics, Chinese Academy of
Sciences, Beijing 100049, P.R. China \\
2) Theoretical Physics Center for Science Facilities, CAS, Beijing
100049, China}

\begin{abstract}

We propose to study the open charm effects in $e^+ e^-\to
J/\psi\eta$, $J/\psi\pi^0$ and $\phi\eta_c$. We show that the
exclusive cross section lineshapes of these processes would be
strongly affected by the open charm effects. Since the final state
light meson productions are through soft gluon radiations, we assume
a recognition of this soft process via charmed meson loops at
hadronic level. A unique feature among these three reactions is that
the $D\bar{D^*}+c.c.$ open channel is located in a relatively
isolated energy, i.e. $\sim 3.876$ GeV, which is sufficiently far
away from the known charmonia $\psi(3770)$ and $\psi(4040)$.
Therefore, the cross section lineshapes of these reactions may
provide an opportunity for singling out the open charm effects with
relatively well-defined charmonium contributions. In particular, we
find that reaction $e^+ e^-\to J/\psi\pi^0$ is sensitive to the open
charm $D\bar{D^*}+c.c.$ Due to the dominance of the isospin-0
component at the charmonium energy region, we predict an enhanced
model-independent cusp effect between the thresholds of
$D^0\bar{D}^0+c.c.$ and $D^+ D^{*-}+c.c.$ This study can also help
us to understand the $X(3900)$ enhancement recently observed by
Belle Collaboration in $e^+ e^-\to D\bar{D}+c.c.$

\end{abstract}

\date{\today}


\maketitle

\section{Introduction}

During the past years the progress in experiment in the study of
hadron spectroscopy has brought a lot of surprises. In the
charmonium sector, a number of new resonance-like signals have been
observed by the B-factories~\cite{Olsen:2009ys}. These observations
have not only initiated tremendous interests in their nature, but
also revived the efforts on the search for exotics in both
experiment and theory (e.g. see
Refs.~\cite{Voloshin:2007dx,Eichten:2007qx,Brambilla:2010cs,Drenska:2010kg}
for a recent review on these issues.

Although various theoretical prescriptions have been proposed in
order to understand the underlying dynamics for the production and
decay of these new ``resonances", such as hybrid charmonium,
tetraquark, baryonium, and hadronic molecule, an interesting feature
with those observed resonance-like signals is that most of them are
close to open charmed meson production thresholds. An example is the
$X(3872)$ which is located in the vicinity of $D^0\bar{D}^{*0}$.
Because of this, a molecular prescription has been broadly
investigated in the literature. An alternative view is its
indication of the underlying non-perturbative mechanisms arising
from open charm thresholds.

Phenomenologically, the open channel effects may have more general
dynamical implications. They would allow different partial waves to
contribute in exclusive processes. In contrast, the hadronic
molecule scenario would require a relative $S$ wave between the
interacting hadrons. On the other hand, an additional $q\bar{q}$
pair creation near the open channel is highly non-perturbative.
Therefore, this non-perturbative mechanism would play an important
role near the open heavy-flavor threshold to shift the hadrons'
masses nearby, and change their wavefunctions and decay properties.
It is realized that a better understanding of the open channel
effects would be a prerequisite for our ultimate understanding of
the hadron spectroscopies.

In this work we shall use an effective Lagrangian approach based on
the heavy quark symmetry and chiral
symmetry~\cite{Li:2008xm,Liu:2009vv,Liu:2010um,Zhang:2009gy,Zhang:2009kr,Zhang:2010zv,Wang:2010iq}
to study the open charm effects in $e^+ e^-\to J/\psi\eta$,
$J/\psi\pi^0$ and $\phi\eta_c$. Our motivation is based on the
following points: i) In  $e^+ e^-\to D\bar{D}+c.c.$, it is observed
by Belle Collaboration~\cite{Pakhlova:2008zza} that an enhancement
around 3.9 GeV, i.e. $X(3900)$, of which the nature is unclear. As
we know, the vector charmonium ($J^{PC}=1^{--}$) spectrum has been
better established since the charmonium states can be directly
produced via the time-like virtual photon in $e^+ e^-$
annihilations. Therefore, the observation of the enhancement
$X(3900)$ provides an ideal place to investigate the underlying
dynamics beyond the known charmonium spectrum. ii) In
Ref.~\cite{Zhang:2009gy}, the $D\bar{D^*}+c.c.$ open charm effects
are investigated and seem to provide a natural explanation for the
$X(3900)$ enhancement without introducing any exotic components. In
order to confirm the nature of the $X(3900)$, one should investigate
other possible reflections of such a mechanism. iii)  Note that the
position of the $X(3900)$ is located between the known $\psi(3770)$
and $\psi(4040)$, its coupling to $J/\psi\eta$ and isospin-violating
$J/\psi\pi^0$ would receive relatively small interferences from the
nearby resonances. Apart from this anticipation, we also consider
the $\phi\eta_c$ channel, of which the threshold is very close to
the $D\bar{D^*}+c.c.$ Therefore, peculiar threshold effects due to
the open $D\bar{D^*}+c.c.$ might be detectable in $e^+ e^-\to
J/\psi\eta$, $J/\psi\pi^0$ and $\phi\eta_c$. Although the
experimental measurement of these three exclusive channels are not
available, the CLEO Collaboration recently provide an upper limit of
the cross sections for $e^+e^-\to J/\psi \eta$ and
$J/\psi\pi^0$~\cite{Coan:2006rv}, which would be a guidance for us
to examine the open charm effects in the vector charmonium
excitations.

This paper is organized as below. In Sec. II we present the
effective Lagrangian approach with formulae. The parameters are
fitted in Sec. III. Section IV is devoted to numerical results and
discussions. The summary is given in the last section.

\section{Formulae}

The effective Lagrangian approach has been successfully applied to
various charmonium decay processes as one of the most important
non-perturbative mechanisms in order to explain some of the
long-standing puzzles in the charmonium energy region. For instance,
it was shown that the open charm coupled-channel effects would lead
to sizeable non-$D\bar{D}$ decay branching ratios for
$\psi(3770)$~\cite{Zhang:2009kr}, and account for the large breaking
of the helicity selection rule in charmonium
decays~\cite{Liu:2009vv,Liu:2010um,Wang:2010iq}.

\begin{figure}[tb]
\begin{center}
\hspace{-7cm}
\includegraphics[scale=0.6]{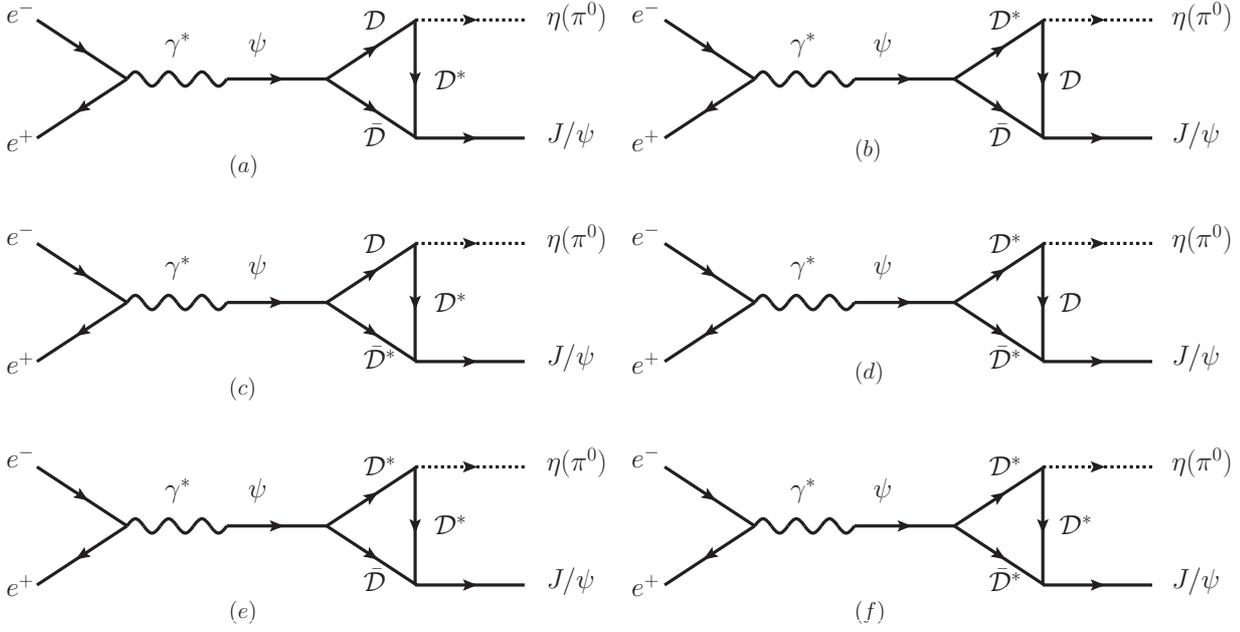}
\vspace{0cm} \caption{Schematic diagrams for $e^+e^-\to
J/\psi\eta(\pi^0)$ via charmed $D$ ($D^*$) meson loops. The diagrams
for the $\phi\eta_c$ mode are similar. }\label{fig:1}
\end{center}
\end{figure}

As we know from the vector meson dominance (VMD) model, light vector
meson contributions to the cross sections are negligible in the
charmonium energy region. The main contributions included here are
from vector charmonium excitations. Note that the final states $VP$
consist of a charmonium plus a light meson. Therefore, the
transitions are Okubo-Zweig-Iizuka (OZI) rule violating processes,
which should be dominated by soft mechanisms near threshold. Since
the pure electromagnetic (EM) transitions are negligibly small, a
natural way to recognize the soft mechanisms for $e^+e^-\to
J/\psi\eta$, $J/\psi\pi^0$ and $\phi\eta_c$ is via the open charm
transitions which are illustrated by Fig. \ref{fig:1}. Similar
approach has been applied to the study of the cross section
lineshape of $e^+e^-\to \omega\pi^0$ in the vicinity of the $\phi$
meson mass region~\cite{Li:2008xm}.

The effective Lagrangians for the coupling vertices involving
charmonia and charmed mesons are extracted from heavy quark
effective theory and chiral symmetry as applied in
Ref.~\cite{Liu:2009vv,Liu:2010um,Wang:2010iq}. They are written as
follows:
\begin{equation}
\mathcal{L}_2=i g_2 Tr[R_{c\bar{c}} \bar{H}_{2i}\gamma^\mu
{\stackrel{\leftrightarrow}{\partial}}_\mu \bar{H}_{1i}] + H.c.,
\end{equation}
where the $S$-wave charmonium states are expressed as
\begin{equation}
R_{c\bar{c}}=\left( \frac{1+ \rlap{/}{v} }{2} \right)\left(\psi^\mu
\gamma_\mu -\eta_c \gamma_5 \right )\left( \frac{1- \rlap{/}{v} }{2}
\right),
\end{equation}
and the charmed and anti-charmed meson triplet are
\begin{eqnarray}
\nonumber H_{1i}&=&\left( \frac{1+ \rlap{/}{v} }{2} \right)[
\mathcal{D}_i^{*\mu} \gamma_\mu -\mathcal{D}_i\gamma_5], \\
H_{2i}&=& [\bar{\mathcal{D}}_i^{*\mu} \gamma_\mu
-\bar{\mathcal{D}}_i\gamma_5]\left( \frac{1- \rlap{/}{v} }{2}
\right), \label{eq:superfield}
\end{eqnarray}
where $\mathcal{D}$ and $\mathcal{D}^*$ are the pseudoscalar charmed
mesons ($(D^{0},D^{+},D_s^{+})$) and vector charmed mesons
($(D^{*0},D^{*+},D_s^{*+})$), respectively. Additionally, the
$J/\psi$ and $\eta_c$, and $\mathcal{D}^*$ and $\mathcal{D}$, can be
considered as doublet states based on the heavy quark spin symmetry.
The Lagrangian describing the interactions between light meson and
charmed mesons reads
\begin{eqnarray}
\mathcal{L}&=&iTr[H_iv^\mu \mathbf{D}_{\mu
ij}\bar{H}_j]+igTr[H_i\gamma_\mu\gamma_5A^\mu_{ij}\bar{H}_j]+i\beta
Tr[H_iv^\mu(V_\mu-\rho_\mu)_{ij}\bar{H}_j]+i\lambda
Tr[H_i\sigma^{\mu\nu}F_{\mu\nu}(\rho)_{ij}\bar{H}_j], \label{vph}
\end{eqnarray}
where the operator
$A_\mu=\frac{1}{2}(\xi^\dag\partial_\mu\xi-\xi\partial_\mu\xi^\dag)$
with $\xi=\sqrt{\Sigma}=e^{\frac{iM}{f_\pi}}$, and
$F_{\mu\nu}(\rho)\equiv
\partial_\mu\rho_\nu-\partial_\nu\rho_\mu+[\rho_\mu,\rho_\nu]$.
$M$ and  $\rho$ denote the light pseudoscalar octet and vector
nonet, respectively~\cite{Cheng:2004ru,Casalbuoni:1996pg},
\begin{eqnarray}
M &=& \left(\matrix{{\pi^0\over\sqrt{2}}+{\eta\over\sqrt{6}} & \pi^+
& K^+ \cr
 \pi^- & -{\pi^0\over\sqrt{2}}+{\eta\over\sqrt{6}} & K^0  \cr
 K^- & \bar{ K^0} & -\sqrt{2\over 3}\eta }\right), \ \
\rho = \left(\matrix{{\rho^0\over\sqrt{2}}+{\omega\over\sqrt{2}} &
\rho^+ & K^{*+} \cr
 \rho^- & -{\rho^0\over\sqrt{2}}+{\omega\over\sqrt{2}} & K^{*0}  \cr
 K^{*-} & \bar{ K^{*0}} & \phi }\right).
 \label{eq:pv}
 \end{eqnarray}
To keep the same convention as Eq.~(\ref{eq:superfield}), the
superfield $H$ is defined as below~\cite{Colangelo:2003sa}:
\begin{eqnarray}
\nonumber
H_i&=&\left(\frac{1+\rlap{/}{v}}{2}\right)[D_i^{*\mu}\gamma_\mu-D_i\gamma_5],\\
\bar{H}_i&=&[D_i^{*\dag\mu}\gamma_\mu+D^\dag_i\gamma_5]\left(\frac{1+\rlap{/}{v}}{2}\right).
\end{eqnarray}
Substituting Eq.~(\ref{eq:superfield}) to Eq.~(\ref{vph}), it is
easy to obtain the detailed form of the interactions to the leading
order~\cite{Cheng:2004ru}:
\begin{eqnarray}
\nonumber
 {\cal L} &=&-g_{\mathcal{D}^*\mathcal{D}\mathcal{P}}(\mathcal{D}^i\partial^\mu
\mathcal{P}_{ij}
 \mathcal{D}_\mu^{*j\dagger}+\mathcal{D}_\mu^{*i}\partial^\mu \mathcal{P}_{ij}\mathcal{D}^{j\dagger})
 +{1\over 2}g_{\mathcal{D}^*\mathcal{D}^*\mathcal{P}}
 \epsilon_{\mu\nu\alpha\beta}\,\mathcal{D}_i^{*\mu}\partial^\nu \mathcal{P}^{ij}
 {\stackrel{\leftrightarrow}{\partial}}{\!^\alpha} \mathcal{D}^{*\beta\dagger}_j
 \\\nonumber
 &-& ig_{\mathcal{D}\mathcal{D}\mathcal{V}} \mathcal{D}_i^\dagger {\stackrel{\leftrightarrow}{\partial}}{\!_\mu} \mathcal{D}^j(V^\mu)^i_j
 -2if_{\mathcal{D}^*\mathcal{D}\mathcal{V}} \epsilon_{\mu\nu\alpha\beta}
 (\partial^\mu \mathcal{V}^\nu)^i_j
 (\mathcal{D}_i^\dagger{\stackrel{\leftrightarrow}{\partial}}{\!^\alpha} \mathcal{D}^{*\beta j}+\mathcal{D}_i^{*\beta\dagger}{\stackrel{\leftrightarrow}{\partial}}{\!^\alpha} D^j)
 \\
 &+& ig_{\mathcal{D}^*\mathcal{D}^*\mathcal{V}} \mathcal{D}^{*\nu\dagger}_i {\stackrel{\leftrightarrow}{\partial}}{\!_\mu} \mathcal{D}^{*j}_\nu(\mathcal{V}^\mu)^i_j
 +4if_{\mathcal{D}^*\mathcal{D}^*\mathcal{V}} \mathcal{D}^{*\dagger}_{i\mu}(\partial^\mu \mathcal{V}^\nu-\partial^\nu
 \mathcal{V}^\mu)^i_j \mathcal{D}^{*j}_\nu,
 \label{eq:LDDV}
 \end{eqnarray}
where $\epsilon_{\alpha\beta\mu\nu}$ is the Levi-Civita tensor and
$D$ meson field destroys a $D$ meson.

The kinematics for a typical transition of Fig.~\ref{fig:1} are
defined as:
$e^+(k_2)e^-(k_1)\to\mathcal{D}(p_1)\bar{\mathcal{D}}(p_2)\left[\mathcal{D}(p_3)\right]\to
J/\psi(k)\eta(q) \ (J/\psi(k)\pi^0(q), \ \phi(k)\eta_c(q))$, with
$k_1$, $k_2$, $k$, and $q$, the four-vector momenta of the
corresponding particles, respectively. Consequently, the transition
amplitude for an intermediate vector charmonium $\psi$ can be
expressed as:
\begin{eqnarray}\label{amp-tot}
 \mathcal{M}&=&\bar{v}(k_2)e\gamma^\mu
u(k_1)\frac{-g_{\mu\nu}}{s}\frac{em_\psi^2}{f_\psi}\frac{-g^{\nu\alpha}+\frac{p^\nu
p^\alpha}{m_\psi^2}}{s-m_\psi^2+im_\psi\Gamma_\psi}\sum_{i=a}^f\mathcal{M}_{i\alpha}
\ ,
\end{eqnarray}
where $\mathcal{M}_{i\alpha}$ could be a complex number serving as a
vertex function for the intermediate $\psi$ coupling to final state
$VP$ via $D$ meson loops. For the processes of Fig.~\ref{fig:1},
$\mathcal{M}_{i\alpha}$ has the following expressions:
\begin{eqnarray}
 \nonumber
\mathcal{M}_{a\alpha}&=&-\int\frac{d^4p_3}{(2\pi)^4}2g_{J/\psi\mathcal{D}\bar{\mathcal{D}}^*}g_{\psi\mathcal{D}\bar{\mathcal{D}}}g_{\mathcal{P}\bar{\mathcal{D}}\mathcal{D}^*}
\epsilon_{\mu\nu\sigma\lambda}p_3^\lambda\epsilon_{J/\psi}^\nu
p_2^\mu q^\sigma
p_{1\alpha}\frac{1}{a_1a_2a_3}\mathcal{F}(p_3^2),\\\nonumber
\mathcal{M}_{b\alpha}&=&-\int\frac{d^4p_3}{(2\pi)^4}2g_{J/\psi\mathcal{D}\bar{\mathcal{D}}}g_{\psi\mathcal{D}\bar{\mathcal{D}}^*}g_{\mathcal{P}\bar{\mathcal{D}}\mathcal{D}^*}
\epsilon_{\mu\alpha\sigma\lambda}p_1^\lambda p_2^\mu q^\sigma
p_2\cdot\epsilon_{J/\psi}\frac{1}{a_1a_2a_3}\mathcal{F}(p_3^2),\\\nonumber
\mathcal{M}_{c\alpha}&=&-\int\frac{d^4p_3}{(2\pi)^4}g_{J/\psi\mathcal{D}^*\bar{\mathcal{D}}^*}g_{\psi\mathcal{D}\bar{\mathcal{D}}^*}g_{\mathcal{P}\bar{\mathcal{D}}\mathcal{D}^*}
\epsilon_{\mu\alpha\rho\lambda}p_2^\lambda p_1^\mu\\\nonumber
&\times&(\epsilon_{J/\psi}^\rho
p_{2\sigma}-p_{3}^\rho\epsilon_{J/\psi\sigma}+2p_3\cdot\epsilon_{J/\psi}g^\rho_{\sigma})
(-g^{\sigma\delta}+\frac{p_3^\sigma
p_3^\delta}{m^2_{\mathcal{D}^*}})q_\delta\frac{1}{a_1a_2a_3}\mathcal{F}(p_3^2),\\\nonumber
\mathcal{M}_{d\alpha}&=&-\int\frac{d^4p_3}{(2\pi)^4}g_{J/\psi\mathcal{D}\bar{\mathcal{D}}^*}g_{\psi\mathcal{D}^*\bar{\mathcal{D}}^*}g_{\mathcal{P}\bar{\mathcal{D}}\mathcal{D}^*}
\epsilon_{\mu\nu\sigma\lambda}p_2^\lambda\epsilon_{J/\psi}^\nu
p_3^\mu (-g^{\delta\iota}+\frac{p_1^\delta
p_1^\iota}{m^2_{\mathcal{D}^*}})\\\nonumber &\times&(2p_{2\alpha}
g_{\delta}^\sigma+p_{1}^\sigma
g_{\delta\alpha}-p_{2\delta}g^\delta_\alpha)
q_\iota\frac{1}{a_1a_2a_3}\mathcal{F}(p_3^2),\\\nonumber
\mathcal{M}_{e\alpha}&=&-\int\frac{d^4p_3}{(2\pi)^4}g_{J/\psi\mathcal{D}\bar{\mathcal{D}}^*}g_{\psi\mathcal{D}\bar{\mathcal{D}}^*}g_{\mathcal{P}\mathcal{D}^*\bar{\mathcal{D}}^*}
\epsilon_{\mu\nu\varsigma\beta}q^\nu p_3^\varsigma\\\nonumber
&\times&\epsilon^{\rho\iota\beta\lambda}p_{3\lambda}\epsilon_{J/\psi\iota}p_{2\rho}
\epsilon^{\tau\alpha\mu\kappa}p_{1\kappa}p_{2\tau}\frac{1}{a_1a_2a_3}\mathcal{F}(p_3^2),\\\nonumber
\mathcal{M}_{f\alpha}&=&\int\frac{d^4p_3}{(2\pi)^4}g_{J/\psi\mathcal{D}^*\bar{\mathcal{D}}^*}g_{\psi\mathcal{D}^*\bar{\mathcal{D}}^*}g_{\mathcal{P}\mathcal{D}^*\bar{\mathcal{D}}^*}
\epsilon_{\mu\nu\varsigma\beta}p_3^\nu p_1^\varsigma\\\nonumber
&\times&(-2p_{2\alpha} g^{\mu\lambda}-p_1^\lambda
g^\mu_\alpha+p_2^\mu
g^\lambda_\alpha)(-g^{\lambda\delta}+\frac{p_2^\lambda
p_2^\delta}{m^2_{\mathcal{D}^*}})\\
&\times&(2p_3\cdot\epsilon_{J/\psi}g_{\delta}^\beta+p_{2}^\beta\epsilon_{J/\psi\delta}-p_{3\delta}\epsilon_{J/\psi}^\beta)\frac{1}{a_1a_2a_3}\mathcal{F}(p_3^2),
\end{eqnarray}
with $a_1\equiv p_1^2-m_1^2$, $a_2\equiv p_2^2-m_2^2$ and $a_3\equiv
p_3^2-m_3^2$. As we know, the meson loop integrals have ultra-violet
divergence. To cut off the unphysical contributions in the high
momentum transfers, we introduce a form factor as broadly applied in
the literature.  A typical dipole form factor for the integrals is
as follows:
\begin{equation}\label{formfactor}
\mathcal{F}(p_3^2)=\left(\frac{\Lambda^2-m_3^2}{\Lambda^2-p_3^2}\right)^2,
\end{equation}
where $\Lambda$ is the cutoff energy and can be parameterized as
$\Lambda=m+\alpha\Lambda_{QCD}$ with $m$ the mass of the exchanged
particle and $\Lambda_{QCD}=220$ MeV.

In this study, we include five resonances, i.e. $J/\psi$,
$\psi(3686)$, $\psi(3770)$, $\psi(4040)$, and $\psi(4160)$. Thus,
the total transition amplitude can be expressed as
\begin{equation}
\mathcal{M}=\mathcal{M}_{J/\psi}+\mathcal{M}_{\psi(3686)}
+e^{i\theta}\mathcal{M}_{\psi(3770)}
+e^{i\beta}\mathcal{M}_{\psi(4040)}+e^{i\phi}\mathcal{M}_{\psi(4160)}
\ ,
\end{equation}
where $\theta$, $\beta$ and $\phi$ are the relative phase angles
which can be determined by experimental data.

In Eq.~(\ref{amp-tot}), the dimensionless vector charmonia couplings
to the virtual photon, ${e}/{f_V}$, can be determined by the VMD
model in $V\to e^+ e^-$:
\begin{eqnarray}
\frac{e}{f_V}&=&\left[\frac{3\Gamma_{V\to
e^+e^-}}{2\alpha_e|\mathbf{p}_e|}\right]^{\frac{1}{2}},
\end{eqnarray}
where $\Gamma_{V\to e^+e^-}$ is the vector meson partial decay width
to $e^+ e^-$, and $\mathbf{p}_e$ is the three-vector momentum of the
final state electron in the vector meson rest frame. With the
partial decay widths from the Particle Data
Group~\cite{Nakamura:2010zzi}, the couplings $e/f_V$ for those
low-lying charmonia are listed in Table~\ref{tab-1}.

\begin{table}[ht]
\caption{The $V\gamma^*$ coupling constant $e/f_V$ determined by
experimental data for $V\to e^+e^-$~\protect\cite{Nakamura:2010zzi}.
} \label{tab-1}
\begin{tabular}{ccc }\hline\hline
 $V\to e^+e^-$ &   Partial decay widths (keV) & $e/f_V$   \\
\hline
$J/\psi\to e^+e^-$ & $5.55 $  &  $2.71\times10^{-2}$  \\
$\psi(3686)\to e^+e^-$ & $2.38$   &  $1.63\times10^{-2}$    \\
$\psi(3770)\to e^+ e^-$   &  $0.26$  &  $5.4\times10^{-3}$
\\
$\psi(4040)\to e^+e^-$ &  $0.86$   &  $9.35\times10^{-3}$  \\
$\psi(4160)\to e^+e^-$ &  $0.83$   &  $9.06\times10^{-3}$  \\
\hline\hline
\end{tabular}
\end{table}

It should be noted that the $V\mathcal{D}^*\bar{\mathcal{D}^*}$
coupling consists of two terms with the  relative angular momentum
$L=1$ between $\mathcal{D}^*$ and $\bar{\mathcal{D}}^*$, i.e.
$\epsilon_{V}\cdot\epsilon_{\bar{\mathcal{D}}^*}\epsilon_{\mathcal{D}^*}\cdot(k-q)
+\epsilon_{\mathcal{D}^*}\cdot\epsilon_{\bar{\mathcal{D}}^*}\epsilon_V\cdot(k-q)$.
The coefficients of these two terms are universal for $J/\psi
\mathcal{D}^*\bar{\mathcal{D}}^*$, i.e. $g_{J/\psi
\mathcal{D}^*\bar{\mathcal{D}}^*}$, while for the $\phi
\mathcal{D}^*\bar{\mathcal{D}}^*$, the coupling structure is
$4f_{\mathcal{D}^*\bar{\mathcal{D}}^*V}(\epsilon_{V}\cdot\epsilon_{\bar{\mathcal{D}}^*}\epsilon_{\mathcal{D}^*}\cdot(k-q)
+\epsilon_{V}\cdot\epsilon_{\mathcal{D}^*}\epsilon_{\bar{\mathcal{D}^*}}\cdot(k-q))
-g_{\mathcal{D}^*\bar{\mathcal{D}}^*V}\epsilon_{\mathcal{D}^*}\cdot\epsilon_{\bar{\mathcal{D}}^*}\epsilon_V\cdot(k-q)$.
Here $k$ and $q$ are the incoming four momentum of $\mathcal{D}^*$
and $\bar{\mathcal{D}}^*$, respectively. The total spin of
$\mathcal{D}^*\bar{\mathcal{D}}^*$ system in the first term is
$S=2$, but $S=0$ in the second term. These two couplings,
$f_{\mathcal{D}^*\bar{\mathcal{D}}^*V}$ and
$g_{\mathcal{D}^*\bar{\mathcal{D}}^*V}$, are equal to each other due
to the heavy quark spin symmetry for a quarkonium coupling to the
charmed mesons.

\section{Parameters}

The charmonium couplings to the charmed mesons are extracted under
the SU(3) flavor symmetry and heavy quark symmetry
\cite{Cheng:2004ru,Casalbuoni:1996pg}:
\begin{eqnarray}\label{eq:hsr}
g_{\psi D\bar{D}^*}=\frac{g_{\psi D\bar{D}}}{\widetilde{M}_D},~~~
g_{\psi D^* \bar{D}^*}=g_{\psi
D\bar{D}^*}\sqrt{\frac{m_{D^*}}{m_D}}m_{D^*},~~~
\widetilde{M}_D=\sqrt{m_Dm_{D^*}},\nonumber \\
g_{\eta_c D\bar{D}^*}=g_{\eta_c
D^*\bar{D}^*}\sqrt{\frac{m_D}{m_{D^*}}}m_{\eta_c}
=2g_2\sqrt{m_{\eta_c}m_Dm_{D^*}},~~~g_2=\frac{\sqrt{m_\psi}}{2m_Df_\psi},
\end{eqnarray}
where $m_\psi$ and $f_\psi$ are the mass and decay constant of
$J/\psi$ with $f_\psi=405$ MeV. Since $J/\psi$ and $\psi(3686)$ are
below the $D\bar{D}$ threshold, their couplings to $D\bar{D}$ cannot
be directly measured by experiment. We adopt $g_{J/\psi
D\bar{D}}=7.44$ which is from the VMD model
\cite{Oh:2007ej,Colangelo:2002mj}. The couplings of
$\psi^\prime$($\psi(3770)$) to $D^{(*)}\bar{D}^{(*)}$ have been
extracted from the cross section lineshape of $e^+e^-\to D\bar{D}$
in Ref.~\cite{Zhang:2009gy}, where quite significant isospin
violation effects are found with the couplings, i.e. $g_{\psi^\prime
D^0\bar{D^0}}=9.05\pm2.34$,
 $g_{\psi^\prime D^+D^-}=7.72\pm1.02$,
$g_{\psi(3770)D^0\bar{D^0}}=13.58\pm1.07$,
$g_{\psi(3770)D^+D^-}=10.71\pm1.75$. Meanwhile, one notices that
these extracted values still possess large uncertainties due to the
relatively poor status of the experimental
data~\cite{Ablikim:2008zz,Ablikim:2008zza,Pakhlova:2008zza}.

In general, such isospin breaking contributions will bring
model-dependence to the predictions for the open charm effects.
Since we are still lacking experimental observables to constrain
these parameters, we assume that to leading order  the couplings
between $\psi^\prime(\psi(3770))$ and the charged and neutral
charmed mesons are the same. Namely, we take the average values for
these couplings, i.e. $g_{\psi^\prime D\bar{D}}=(9.05+7.72)/2=8.4$,
and $g_{\psi(3770)D\bar{D}}=(13.58+10.71)/2=12.1$. By requiring that
the cross sections of $e^+e^-\to \psi^\prime\to J/\psi\eta$ agree
with the experimental data, we can determine the form factor
parameter $\alpha$, which can be then fixed in the predictions for
$e^+e^-\to J/\psi\pi^0$ and $\phi\eta_c$. We also take the average
value of charged and neutral ones for strange-charmed mesons. The
other couplings between charmonium and charmed mesons can be
obtained from Eq.~(\ref{eq:hsr}).

To evaluate the couplings of the $\psi(4040)$ to the charmed mesons,
we assume that the phase space allowed
$\mathcal{D}\bar{\mathcal{D}}$,
$\mathcal{D}^*\bar{\mathcal{D}}+\mathcal{D}\bar{\mathcal{D}}^*$ and
$\mathcal{D}^*\bar{\mathcal{D}}^*$ modes account for the total width
of $\psi(4040)$. BaBar Collaboration measured the branching ratio
fractions of these channels~\cite{:2009xs}, i.e. $Br(\psi(4040\to
D\bar{D}))/Br(\psi(4040)\to D^*\bar{D})=0.24\pm0.05\pm0.12$ and
$Br(\psi(4040\to D^*\bar{D}^*))/Br(\psi(4040)\to
D^*\bar{D})=0.18\pm0.14\pm0.03$. The center values are used here to
extract the coupling $g_{\psi(4040)D\bar{D}}=2.02$,
$g_{\psi(4040)D^*\bar{D}^*}=4.24$ and $g_{\psi(4040)D\bar{D}^*}=1.6$
GeV$^{-1}$. The couplings of the $\psi(4160)$ to the charmed mesons
can be obtained in the same way using the data, $Br(\psi(4160)\to
D\bar{D})/Br(\psi(4160)\to D^*\bar{D}^*)=0.02$ and $Br(\psi(4160)\to
D^*\bar{D})/Br(\psi(4160)\to D^*\bar{D}^*)=0.34$ from
Ref.~\cite{:2009xs}. It gives $g_{\psi(4160)D\bar{D}}=0.53$,
$g_{\psi(4160)D^*\bar{D}}=0.71$ GeV$^{-1}$ and
$g_{\psi(4160)D^*\bar{D}^*}=3.08$. This can be regarded as a
reasonable way to extract the couplings of $\psi(4040)$ and
$\psi(4160)$.

For the light meson couplings to the charmed mesons, they are
determined as those in Refs.~\cite{Cheng:2004ru,Zhang:2009gy}:
\begin{eqnarray}
g_{PD\bar{D}^*}=\frac{2g}{f_\pi}\sqrt{m_Dm_{D^*}},~~~g_{PD^*\bar{D}^*}=\frac{g_{PD\bar{D}^*}}{\sqrt{m_Dm_{D}^*}},\\
g_{DDV}=g_{D^*\bar{D}^*V}=\frac{\beta
g_V}{\sqrt{2}},~~~g_{D\bar{D}^*V}=\frac{f_{D^*\bar{D}^*V}}{m_{D^*}}=\frac{\lambda
g_V}{\sqrt{2}},~~~g_V=\frac{m_\rho}{f_\pi}
\end{eqnarray}
where $g=0.59$, $\beta=0.9$, $\lambda=0.56$ GeV$^{-1}$ and
$f_\pi=132$ MeV. Since the SU(3) flavor symmetry works well at
leading order in this energy scale, we adopt the following
relations:
 $g_{D^0\bar{D}^0 (u\bar{u})}=g_{D^+D^-(d\bar{d})}=g_{D_s^+D_s^-(s\bar{s})}$
and $g_{D\bar{D}(s\bar{s})}=g_{D_s\bar{D_s}(q\bar{q})}=0$, where
$q\bar{q}$ stands for a non-strange light quark-antiquark pair. For
the pion coupling, $g_{\pi D
 \bar{D}^*}=\sqrt{2}g_{D\bar{D}^*(q\bar{q})(0^-)}$ is employed.

The flavor wavefunctions of $\eta$ and $\eta^\prime$ are as below,
\begin{eqnarray}
 \eta&=&\cos\alpha_p|n\bar{n}\rangle-\sin\alpha_p|s\bar{s}\rangle,\\
 \eta^\prime&=&\sin\alpha_p|n\bar{n}\rangle+\cos\alpha_p|s\bar{s}\rangle,
 \end{eqnarray}
 where $|n\bar{n}\rangle\equiv |u\bar{u}+d\bar{d}\rangle/\sqrt{2}$ and
 $\alpha_p\equiv \theta_p+\arctan\sqrt{2}$ with
 $\theta_p=-19.1^\circ$~\cite{Zhang:2009gy}.

\section{Numerical results and discussions}

In this Section we present the calculated cross sections for $e^+
e^-\to J/\psi\eta, \ J/\psi\pi^0$, and $\phi\eta_c$ in terms of the
c.m. energy $W$. Five charmonium states are included, i.e. $J/\psi$,
$\psi(3686)$, $\psi(3770)$, $\psi(4040)$ and $\psi(4160)$, which are
the main resonance contributions in the energy region that we are
interested in. As pointed out in the Introduction, these processes
are highly non-perturbative near threshold, which gives rise to the
contributions from the vector charmonia via the charmed meson loops
as a natural mechanism to evade the OZI rule.

\begin{figure}[tb]
\begin{center}
\includegraphics[scale=0.5]{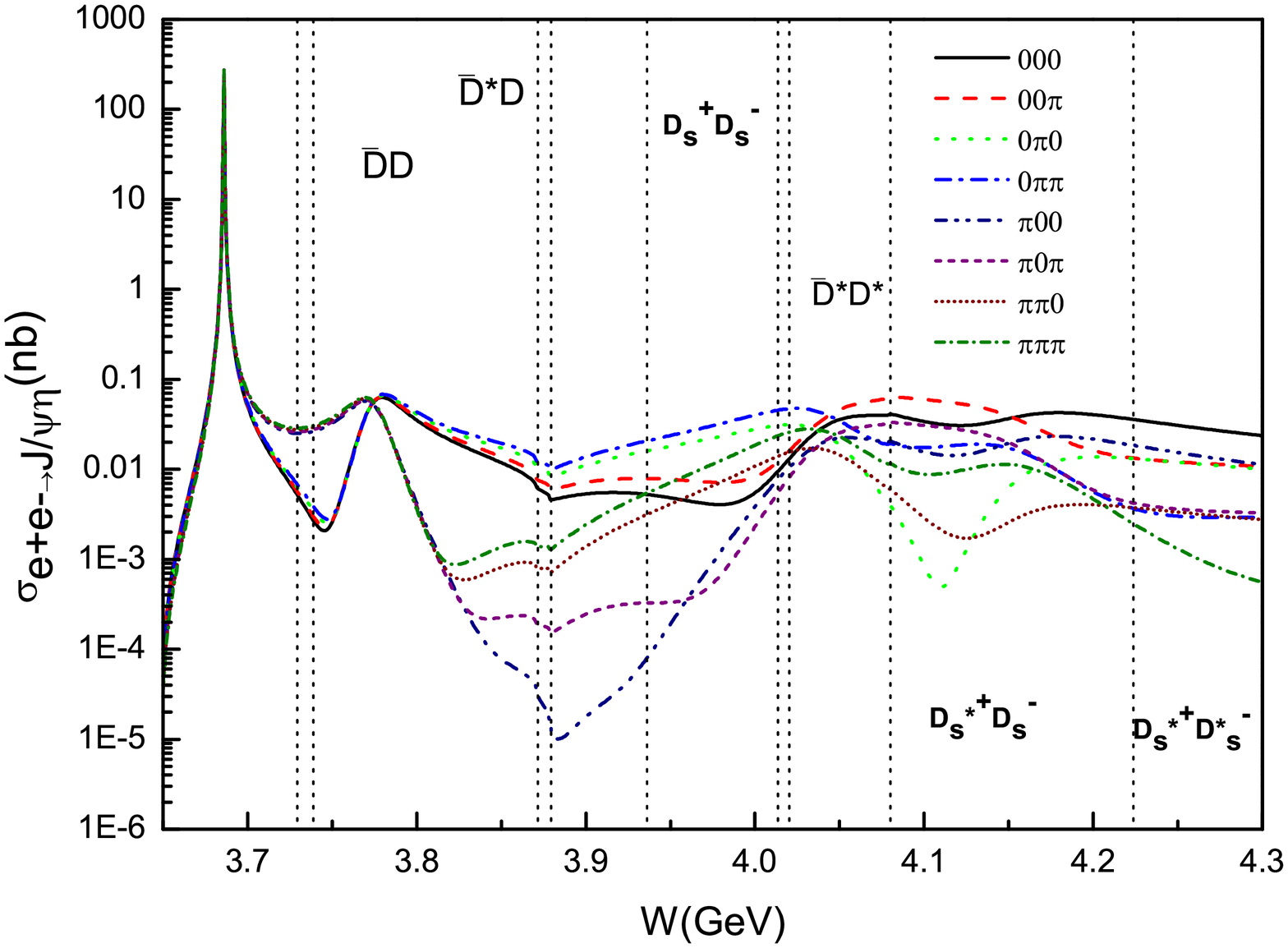}
\caption{The predicted cross section for  $e^+e^-\to J/\psi\eta$ in
terms of the c.m. energy $W$ with the  cutoff parameter
$\alpha=1.57$. The cross sections with different phases, i.e.
$(\theta,
\beta,\phi)=(0, 0,0), \ (0,0,\pi), \ (0,\pi,0), \ (0,\pi,
 \pi), \ (\pi, 0,0), \ (\pi,0,\pi), \ (\pi,\pi,0), \ (\pi,\pi,
 \pi)$, are presented and denoted by different curves. The vertical lines labels the open charm thresholds.
}\label{fig:2}
\end{center}
\end{figure}

\begin{figure}[tb]
\begin{center}
\includegraphics[scale=0.5]{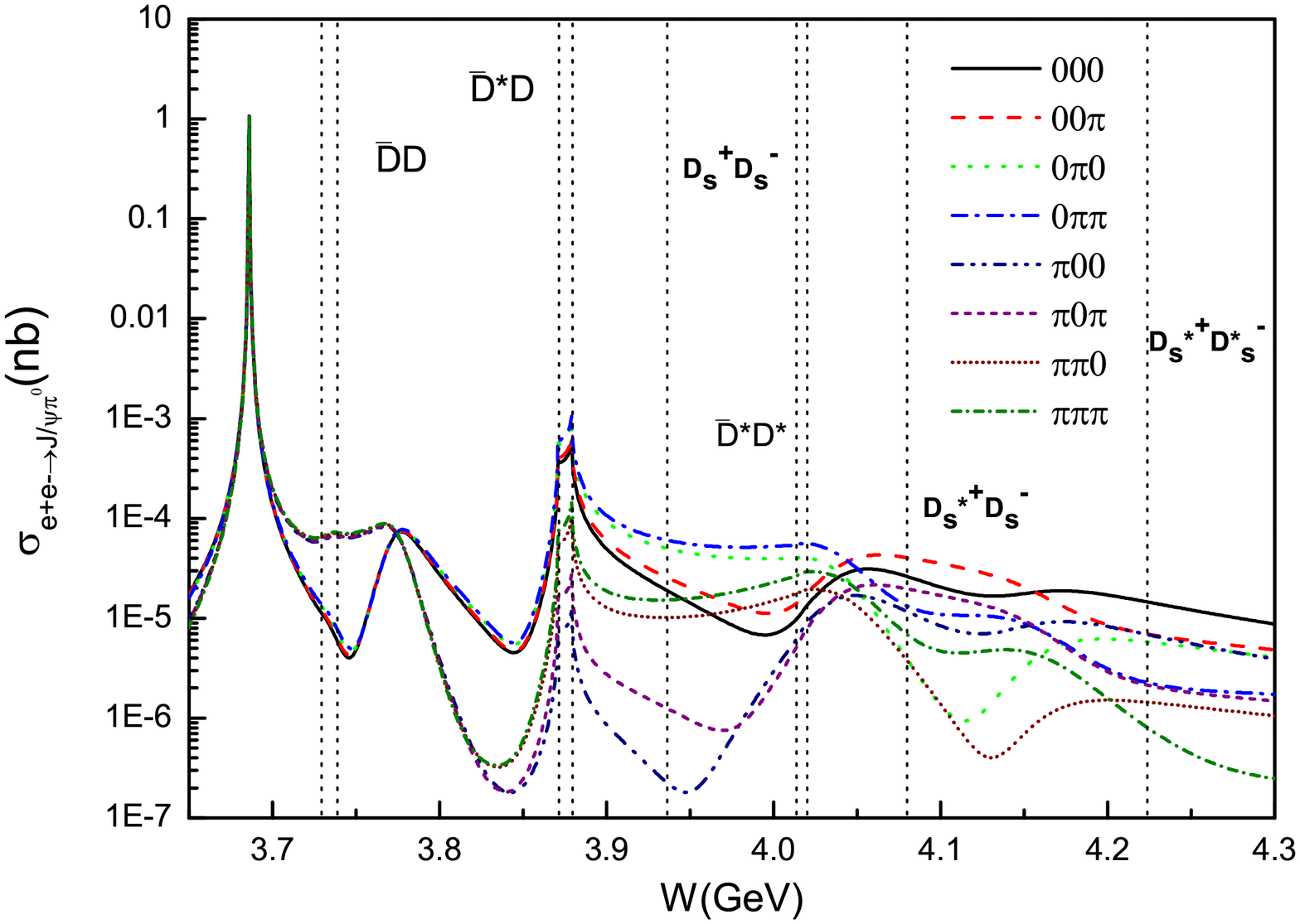}
\caption{The predicted cross section for $e^+e^-\to J/\psi\pi^0$ in
terms of the c.m. energy $W$. The notations are similar to
Fig.~\protect\ref{fig:2}. }\label{fig:3}
\end{center}
\end{figure}

\begin{figure}[tb]
\begin{center}
\includegraphics[scale=0.5]{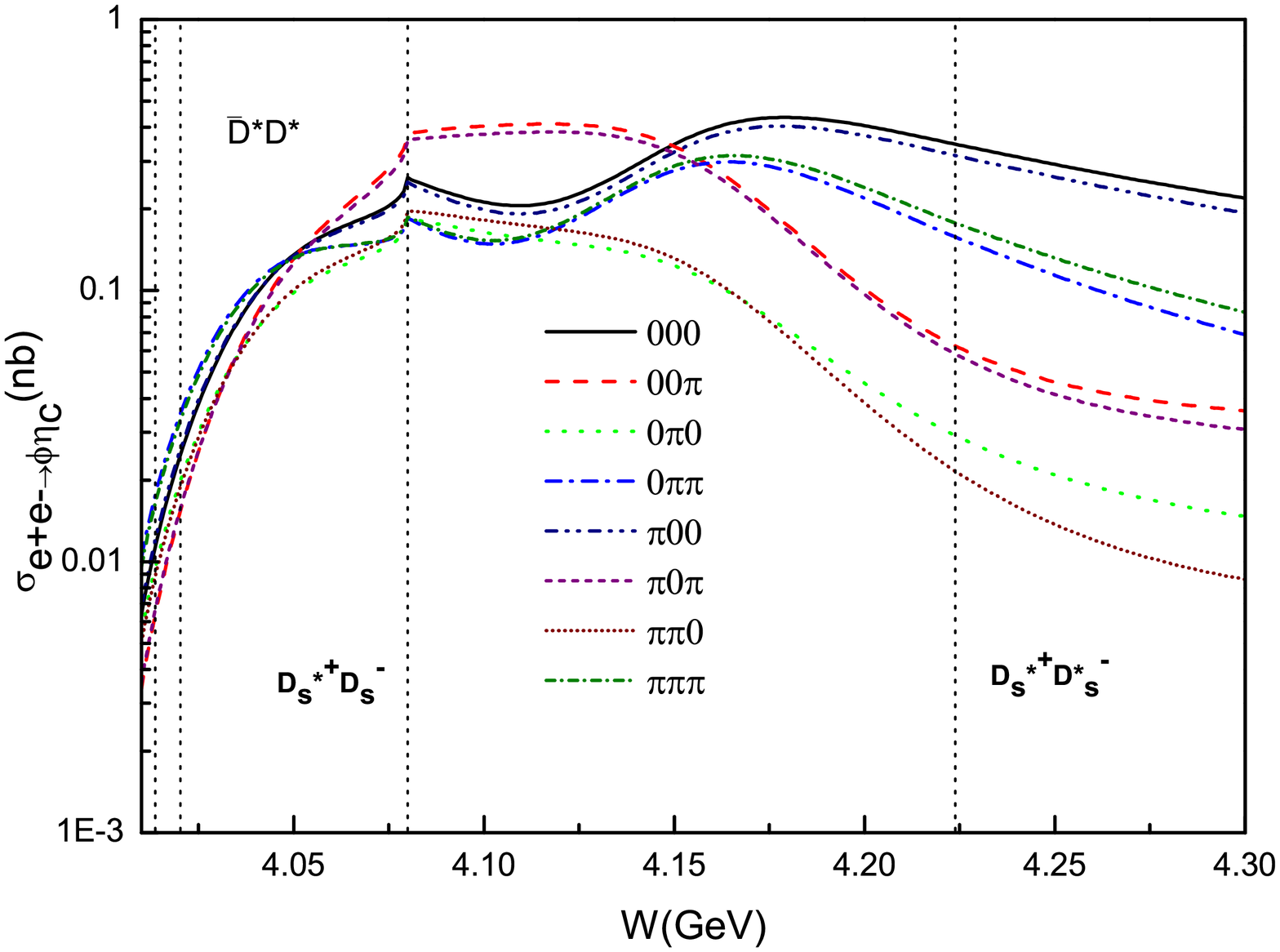}
\caption{The predicted cross section for $e^+e^-\to \phi\eta_c$ in
terms of the c.m. energy $W$.  The notations are similar to
Fig.~\protect\ref{fig:2}. }\label{fig:4}
\end{center}
\end{figure}

Since there are no data available to constrain the relative phases
among the resonance transition amplitudes, we shall examine several
phase combinations to test the sensitivities of the cross sections
to the relative phases. Since the contribution from $J/\psi$ is
negligibly small, we simply take it in phase with the $\psi(3686)$.
The other resonance amplitudes can shift phases in respect of the
$\psi(3686)$. Apart from the phase angles, the only parameter left
is the form factor parameter $\alpha=1.57$, which is fixed by the
cross section $\sigma(e^+e^-\to \psi^\prime\to
J/\psi\eta)=8351.5\times 3.28\%=274 \ nb$ at the mass of
$\psi^\prime$, with the $\psi^\prime$ production cross section
$\sigma(e^+e^-\to \psi^\prime)=8351.5 \ nb$, and $BR(\psi^\prime\to
J/\psi\eta)=3.28\%$~\cite{Nakamura:2010zzi}. As broadly applied in
the literature, the form factor should cut off the unphysical
contributions in the region sufficiently far away from the
singularity. We shall discuss later that the introduction of form
factors may cause unphysical thresholds which should be
distinguished from the physical ones. With the other coupling
parameters fixed in the previous Section, the calculated cross
sections in the first scheme are presented in Figs.~\ref{fig:2},
\ref{fig:3} and \ref{fig:4} for $e^+e^-\to J/\psi\eta$,
$J/\psi\pi^0$ and $\phi\eta_c$, respectively.

Our main results are summarized as follows:

i) The dominant contributions are from the $\psi(3686)$ in
$e^+e^-\to J/\psi\eta$ and $J/\psi\pi^0$. Although the $\psi(3686)$
is below the $\phi\eta_c$ threshold, it still plays an important
role in $e^+e^-\to \phi\eta_c$.  It is because the mass of the
$\psi(3686)$ is close to the energy region considered here, and the
coupling constant of $\psi(3686)$ to the virtual photon is two times
larger than that of the $\psi(3770)$. In contrast, although the
coupling constants of $\psi(4040)$ and $\psi(4160)$ to the virtual
photon are compatible with that of the $\psi(3770)$, their couplings
to the charmed mesons are rather small. Thus, they give relatively
small contributions to the cross sections.

ii) As shown by Figs.~\ref{fig:2} and \ref{fig:3}, the lineshape of
the $\psi(3770)$ is shifted significantly by the $D\bar{D}^*$
threshold. It is an evidence that the open charm coupled-channel
effects would shift the lineshape of the particle nearby. The same
phenomenon appears in $e^+e^-\to\phi\eta_c$ process at the energy
about $4.08$ GeV due to the $D_s^+D_s^{*-}$ threshold as illustrated
in Fig.~\ref{fig:4}. It can be understood, when amplitudes of
different meson loops are added together, it would make the vector
charmonium contributions non-trivial. In particular, since masses of
the thresholds of the open charms are different, the open charm
effects that distort the Breit-Wigner would then be highlighted.

iii) In Figs.~\ref{fig:2} and \ref{fig:3}, the open
$D^{(*)}\bar{D}^{(*)}$ thresholds are explicitly denoted. As
mentioned earlier, the introduction of form factors may cause
unphysical thresholds in the cross section lineshape. Thus, it is
necessary to clarify this in order to correctly understand the
calculated results.

It shows that the dipole form factor of
Eq.~(\ref{formfactor}) should be more suitable for the study of
cross section lineshape and would not introduce additional
thresholds apart from $m_1+m_2$. This can be seen from the
regularization of the propagators in association with the dipole
form factor:
\begin{eqnarray}
&&\frac{1}{p_1^2-m_1^2}\frac{1}{p_2^2-m_2^2}\frac{1}{p_3^2-m_3^2}
\left(\frac{\Lambda_3^2-m_3^2}{\Lambda_3^2-p_3^2}\right)^2\nonumber\\
&\sim &
C(s,m_v^2,m_p^2,m_1^2,m_2^2,m_3^2)-C(s,m_v^2,m_p^2,m_1^2,m_2^2,\Lambda_3^2)\nonumber\\
&&+\frac{\Lambda_3^2-m_3^2}{\varepsilon}\left[C(s,m_v^2,m_p^2,m_1^2,m_2^2,\Lambda_3^2+\varepsilon)
-C(s,m_v^2,m_p^2,m_1^2,m_2^2,\Lambda_3^2)\right], \label{cdipole}
\end{eqnarray}
where function $C$ is the three-point function, and $\varepsilon$ is
a small quantity.

As a comparison, a tri-monopole form factor will cause unphysical
thresholds, namely,
\begin{eqnarray}
\mathcal{F}(p_i^2)\equiv
\prod^3_{i=1}\left(\frac{\Lambda_i^2-m_i^2}{\Lambda_i^2-p_i^2}\right)
\ , \label{tripole}
\end{eqnarray}
where $m_i \ (p_i)$ is the mass (four momentum) of the exchanged
particle, and $\Lambda_i\equiv m_i+\alpha\Lambda_{QCD}$. In this
case, the regularization leads to unphysical thresholds,
$m_1+\Lambda_2$, $\Lambda_1+m_2$ and $\Lambda_1+\Lambda_2$, in the
cross section. This reflects the model-dependent feature arising
from the form factors. In particular, we point out that the cusp
effects caused by these unphysical thresholds would be amplified in
$e^+e^-\to J/\psi\pi^0$, although their effects are negligibly small
in $e^+e^-\to J/\psi\eta$ and $\phi\eta_c$.

iv) The above analysis helps us to identify model-independent
features produced by open charm thresholds. We stress that the
isospin violating transitions in $e^+e^-\to J/\psi\pi^0$ would
provide a great opportunity for disentangling the open charm
effects. Comparing the results of Figs.~\ref{fig:2} and \ref{fig:3},
we can see that the predicted cross sections for $J/\psi\pi^0$ are
greatly suppressed. For $e^+e^-\to J/\psi\pi^0$, since the
contributing intermediate vector charmonia are mainly from $\psi$
resonances with isospin 0, the cross sections would have vanished if
isospin symmetry were conserved. In Fig.~\ref{fig:3}, the
non-vanishing cross sections are produced by the mass differences
(as a result of isospin violation) between the charged and
charge-neutral $D$ ($D^*$) mesons in the intermediate meson loops.
Namely, the charged and charge-neutral meson loop amplitudes cannot
cancel out completely. As a consequence, a peak (cusp) appears
between the thresholds of $D^0\bar{D}^{*0}+c.c.$ and
$D^+D^{*-}+c.c.$ which stands like a resonance, i.e. so-called
$X(3900)$ around 3.876 GeV.

Although the cross sections for both $e^+e^-\to\psi\to J/\psi\eta$
and $J/\psi\pi^0$ are rather sensitive to the relative phases
introduced among the transition amplitudes, the peak structure $X(3900)$
has a model-independent feature and can be searched in experiment.
More importantly, since the thresholds of
$D^0\bar{D}^{*0}+c.c.$ and $D^+D^{*-}+c.c.$ are isolated from the
known $\psi(3770)$ and $\psi(4040)$, the enhancement here would be a clear
evidence for non-resonant peaks in $e^+e^-$ annihilations. In
contrast, although the $D\bar{D}^*+c.c.$ loops have relatively large
contributions to the cross sections in $e^+e^-\to J/\psi\eta$, their
contributions are submerged by other amplitudes and cannot be
indisputably identified in the cross section lineshape. In this sense, the
observation of the $X(3900)$ by the Belle
Collaboration~\cite{Pakhlova:2008zza} may have suggested a hint of the open
$D\bar{D}^*+c.c.$ effects in $e^+e^-\to D\bar{D}$,
but should be further investigated in the $J/\psi\pi^0$ channel.

We also note that  the $\psi(3686)$ has a predominant contribution to
$e^+e^-\to\psi\to J/\psi\pi^0$  due to its strong
isospin violation couplings via the $D$ meson loops~\cite{Guo:2009wr,Guo:2010ak}.
Such a resonance enhancement should be detectable of which the cross section measurement
will provide a calibration for the $X(3900)$ structure.

v) It should be pointed out that this
structure as the open charm effect is a collective one from the
$D\bar{D}^*+c.c.$ loops to which all the vector charmonia have
contributions. That is why such a $P$ wave configuration between
$D\bar{D}^*+c.c.$ can produce the significant enhancement in
 $e^+e^-\to J/\psi\pi^0$. This mechanism is much likely to be different
from the $X(3872)$, which has been broadly investigated in the
literature as a dynamically generated $D\bar{D}^*+c.c.$ bound state
in a relative $S$ wave.

vi) It is interesting to see the model predictions for  $e^+e^-\to
\phi\eta_c$ in Fig.~\ref{fig:4}. In this case, the physical open
charm threshold is $D_s^{*+}D_s^- +c.c.$ which causes rather
significant cusp effects in the cross section. Since the cusp is
close to the $\psi(4040)$ mass, interferences between the open
$D_s^{*+}D_s^- +c.c.$ and $\psi(4040)$ can be investigated. In
addition, although the cross sections exhibit obvious dependence on
the relative phases, we can still see some systematic trends in
terms of the c.m. $W$.

vii) We emphasize again that the relative phases among the
amplitudes would lead to very different predictions for the cross
section lineshapes as illustrated by those curves in
Figs.~\ref{fig:2}, \ref{fig:3} and \ref{fig:4}. Because of this, it
is important to have experimental constraints for the model
parameters. As mentioned earlier, there are measurements of the
cross sections from the CLEO Collaboration at several
energies~\cite{Coan:2006rv}. Although only the upper limits of the
cross sections are provided, it can still give a rough guidance for
the parameter ranges adopted in the calculations. As an example, we
compare the experimental upper limits with the predicted cross
sections with phases $(\theta, \beta,\phi)=(0, 0, 0)$ and $(\pi, 0,
0)$ in Tables~\ref{table-eta} and \ref{table-pi} for $e^+e^-\to
J/\psi\eta$, $J/\psi\pi^0$, respectively. It shows that the
predicted cross sections with the adopted parameters are consistent
with the so-far available experimental information, and indeed give
the correct orders of magnitude of the cross sections.

\begin{center}
\begin{table}
\caption{Comparison of the cross section of $e^+e^-\to J/\psi\eta$
between the experiment data from CLEO~\cite{Coan:2006rv} and our
results with phase $(\theta, \beta,\phi)=(0, 0, 0)$ and $(\pi,0,0)$.
The form factor parameter $\alpha=1.57$ is adopted in the
calculation.} \label{table-eta}
\begin{tabular}{cccc}
  \hline\hline
  $\sigma(e^+e^-\to J/\psi\eta)$ & $3.97\sim4.06$ GeV & $4.12\sim4.2$ GeV & $4.26$ GeV \\
  \hline
  CLEO~\cite{Coan:2006rv} & $<29 \ pb$ & $15^{+5}_{-4}\pm 8 \ pb$ & $<32 \ pb$ \\
 \hline
 Results with $(0,0,0)$& $3.8\sim 39 \ pb$ & $28.9\sim 42.9 \ pb$ & $27.6 \ pb$\\
  \hline
 Results with $(\pi,0,0)$ & $0.57\sim 22.2 \ pb$ & $14.9\sim 24.2 \ pb$ & $14.9 \ pb$ \\
  \hline\hline
\end{tabular}
\end{table}
\end{center}

\begin{center}
\begin{table}
\caption{Comparison of the cross section of $e^+e^-\to J/\psi\pi$
between the experiment data from CLEO~\cite{Coan:2006rv} and our
results with phase $(\theta, \beta,\phi)=(0, 0, 0)$ and $(\pi,0,0)$.
The form factor parameter $\alpha=1.57$ is adopted.}
\label{table-pi}
\begin{tabular}{ccccc}
  \hline\hline
  $\sigma(e^+e^-\to J/\psi\pi^0)$  & $3.97\sim4.06$ GeV & $4.12\sim 4.2$ GeV & $4.26$ GeV \\
  \hline
  CLEO~\cite{Coan:2006rv} &  $<10 \ pb$ & $<3 \ pb$ & $<12 \ pb$ \\
  \hline
  Results with $(0,0,0)$ & $(6.67\sim 28.3)\times 10^{-3} \
  pb$ & $(16.4 \sim 20.0)\times 10^{-3} \ pb $ & $1.2\times 10^{-2} \ pb  $\\
  \hline
  Results with $(\pi,0,0)$ &  ~~$(0.45\sim 17)\times 10^{-3} \ pb$ & ~~$(6.92\sim 9.1)\times 10^{-3} \ pb$ & ~~$5.07\times 10^{-3} \ pb$ \\
  \hline\hline
\end{tabular}
\end{table}
\end{center}


\section{Summary}

In summary, we have proposed to study the coupled channel effects in
$e^+e^-$ annihilating to $J/\psi\eta$, $J/\psi\pi^0$ and
$\phi\eta_c$.  In particular, we show that the reaction $e^+e^-\to
J/\psi\pi^0$ will be extremely interesting for disentangling the
resonance contributions and open charm effects taking the advantage
that the open $D\bar{D^*}$ threshold is relatively isolated from the
nearby known charmonia $\psi(3770)$ and $\psi(4040)$. Although we
also find that the predicted cross sections are rather sensitive to
the model parameters adopted, we clarify that the open charm effects
from the $D\bar{D^*}+c.c.$ channel are rather model-independent.
Therefore, it is extremely interesting to search for the predicted
enhancement around 3.876 GeV (i.e. $X(3900)$) in experiment.
Confirmation of this prediction would allow us to learn a lot about
the nature of non-pQCD in the charmonium energy region.

\section*{Acknowledgments}

The authors thank useful discussions with Changzheng Yuan and
Jingzhi Zhang during revising the manuscript. This work is
supported, in part, by the National Natural Science Foundation of
China (Grants No. 11035006), Chinese Academy of Sciences
(KJCX2-EW-N01), and Ministry of Science and Technology of China
(2009CB825200).

\end{document}